\documentclass[pdflatex,sn-mathphys-num]{sn-jnl}

\usepackage{graphicx}%
\usepackage{multirow}%
\usepackage{amsmath,amssymb,amsfonts}%
\usepackage{amsthm}%
\usepackage[title]{appendix}%
\usepackage{xcolor}%
\usepackage{textcomp}%
\usepackage{manyfoot}%
\usepackage{booktabs}%
\usepackage{algorithm}%
\usepackage{algorithmicx}%
\usepackage{algpseudocode}%
\usepackage{listings}%
\usepackage{siunitx}
\DeclareSIUnit\angstrom{\text{\AA}}
\usepackage{url}
\usepackage{adjustbox}
\usepackage{float}
 \usepackage{array}
 \usepackage{placeins}
 \newcolumntype{L}[1]{>{\raggedright\arraybackslash}p{#1}}
 \newcolumntype{C}[1]{>{\centering\arraybackslash}p{#1}}

\theoremstyle{thmstyleone}%

\theoremstyle{thmstyletwo}%

\theoremstyle{thmstylethree}%

\begin{document}

\title{Simultaneous Fragment Docking for Geometrically Linkable Pose Pairs}

\author[1]{\fnm{Jiyun} \sur{Lee}}
\author[1]{\fnm{You Kyoung} \sur{Chung}}
\author*[1,2]{\fnm{Joonsuk} \sur{Huh}}\email{joonsukhuh@yonsei.ac.kr}

\affil[1]{\orgdiv{Department of Chemistry}, \orgname{Yonsei University},
  \orgaddress{\city{Seoul}, \postcode{03722}, \country{Republic of Korea}}}

\affil[2]{\orgdiv{Department of Quantum Information}, \orgname{Yonsei University},
  \orgaddress{\city{Incheon}, \postcode{21983}, \country{Republic of Korea}}}

\abstract{Computational molecular design requires binding arrangements that are not only energetically favorable but also chemically realizable. However, computational methods remain limited in directly recovering fragment pose pairs that can later be connected into a single molecule. To address this problem, we formulated the simultaneous placement of two fragments as a quadratic unconstrained binary optimization problem, Q--SFD, and introduced an explicit inter-fragment distance term to favor reconstruction-feasible arrangements. Relative to the formulation without this term, Q--SFD approximately doubled top-1 recovery of reconstruction-feasible pairs, and the top-5 solutions contained at least one feasible pair for more than 90\% of benchmark cases without loss of fragment-level pose accuracy.}


\maketitle

\section{Introduction}

Computational drug discovery relies heavily on protein--ligand interaction energy calculations because they directly describe binding behavior. In structure-based drug design (SBDD), these calculations are commonly used to evaluate how a ligand may bind within a protein binding site and to rank plausible binding poses. However, accurate pose prediction becomes increasingly difficult as ligand flexibility increases, because the conformational search space grows rapidly with the number of internal degrees of freedom.

Fragment-based drug discovery (FBDD) offers a practical way to manage this complexity. Instead of screening large and flexible compounds, it starts from small fragments with low structural complexity. Although individual fragments usually bind weakly, their high ligand efficiency makes them useful starting points for optimization~\cite{rees2004fragment,erlanson2016twenty,murray2009rise}. Through iterative structural and medicinal-chemistry refinement, these fragments can be developed into higher-affinity lead compounds. Because they explore chemical space efficiently at a manageable molecular size, FBDD has become a widely used strategy in early-stage drug discovery. Common fragment-to-lead routes include fragment growing, merging, and linking~\cite{lamoree2017current,deSouzaNeto2020insilico}.

Among these routes, fragment linking is particularly attractive. In principle, two fragments that bind in nearby regions of the same binding site can be connected into a single ligand with improved affinity~\cite{bancet2020fragment,murray2002entropy}. At the same time, linking remains one of the most difficult fragment-to-lead strategies in practice. Success depends not only on the binding of the individual fragments, but also on linker geometry, preservation of the original poses after covalent connection, internal strain in the linked molecule, and the entropic cost of combining two binding elements into one ligand~\cite{jencks1981additivity,murray2002entropy,kirsch2019concepts,ichihara2011compound}. When these factors are unfavorable, linking often fails to produce the expected gain in potency~\cite{mondal2016fragment,bancet2020fragment}.

For this reason, a useful computational result for linking is not simply a pair of poses that appears reasonable when each fragment is considered separately. It should remain compatible with reconstruction into a physically reasonable single molecule. Conventional independent docking workflows do not directly address this requirement. When two fragments are docked separately, the top-ranked pose of each fragment may look plausible on its own while still being incompatible with the other. The two fragments may sterically overlap, point in unfavorable connection directions, or place their attachment atoms too far apart for chemically reasonable linking~\cite{hartshorn2005fragment}. These limitations are especially important in early fragment-based discovery, where fragment binding is often weak and docking scores alone may not reliably distinguish true binding modes from nonspecific associations~\cite{murray2009rise,hajduk2007decade}.

Several existing approaches address parts of this problem. Independent fragment docking can be followed by linker design or geometric filtering~\cite{deSouzaNeto2020insilico,bissaro2020revisiting}. Other methods focus on stepwise ligand construction or on placing multiple ligands within the same binding site~\cite{rarey1996fast,allen2015dock,eberhardt2021_vina120}. These approaches are useful for pose exploration, ligand construction, or simultaneous placement. However, they do not directly optimize for placements that remain suitable for later covalent connection. In most workflows, linkability is checked only after candidate poses have already been generated.

In this study, we focus on a narrower but practically important task. We aim to recover a pose pair for two rigid fragments that are intended to be connected into a single ligand. We refer to this output as a \emph{reconstruction-feasible pose pair}. In such a pair, the two attachment sites must be separated by a chemically plausible distance, and the overall arrangement must remain physically consistent after reconstruction~\cite{bancet2020fragment,mondal2016fragment}. This requirement becomes especially important when no crystal structure captures both fragments together in the same binding site, which is a common situation in early fragment-based discovery campaigns~\cite{hajduk2007decade,deSouzaNeto2020insilico}.

To address this problem, we formulate the simultaneous placement of two rigid fragments as a quadratic unconstrained binary optimization (QUBO) problem and refer to the resulting framework as QUBO-based simultaneous fragment docking (Q-SFD). In this framework, placements are selected on a shared interaction grid using binary decision variables. The objective function combines fragment--protein interaction terms, rigid-body consistency, and an explicit inter-fragment distance term that favors attachment-site separations compatible with later covalent connection. This representation is well suited to the problem because placement is fundamentally a discrete selection problem, and geometric relationships can be expressed through pairwise interactions.

\begin{figure*}[t]
  \centering
  \includegraphics[width=\textwidth]{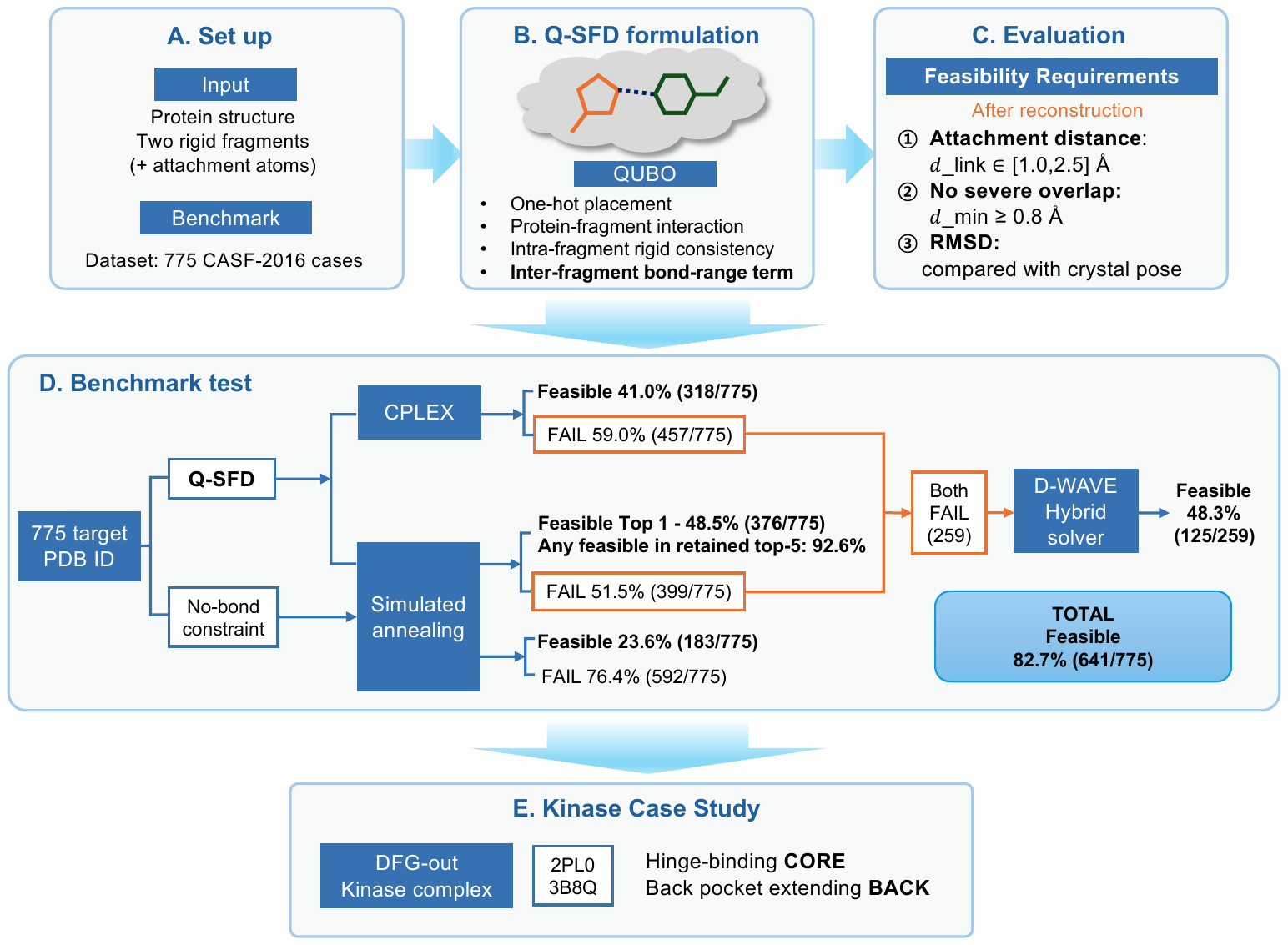}
  \caption{Overview of the Q-SFD study design and workflow. The figure summarizes the overall pipeline from benchmark setup to formulation, evaluation, and downstream analyses. (A) Input preparation and retrospective benchmark construction from 775 CASF-2016-derived cases. (B) Q-SFD formulation for simultaneous placement of two rigid fragments, combining one-hot placement, protein--fragment interaction, intra-fragment rigid consistency, and an inter-fragment distance term. (C) Post-reconstruction evaluation criteria based on attachment distance, severe overlap, and RMSD to the crystallographic reference. (D) Main benchmark comparisons among Q-SFD, the version without the inter-fragment distance term, CPLEX, and a supplementary hybrid follow-up on the stringent subset that failed under both simulated annealing and CPLEX. (E) Domain-guided kinase case study illustrating practical deployment of the framework on representative Asp-Phe-Gly-out (DFG-out) kinase complexes.}
  \label{fig:study_overview}
\end{figure*}

We evaluated Q-SFD on a retrospective benchmark of 775 cases derived from crystallographic ligands. We compared the formulation that included the inter-fragment distance term with a version that did not include this term, as well as with IBM CPLEX as an exact solver applied to the same QUBO. We assessed the results using reconstruction feasibility, attachment-distance distributions, and fragment-level root-mean-square deviation (RMSD). Supplementary backend analyses and a domain-guided kinase case study were used to contextualize solver behavior and practical deployment under the same formulation. The overall study design and analysis workflow are summarized in Fig.~\ref{fig:study_overview}.

\section{Results}

\subsection{The inter-fragment distance term improves top-1 recovery of reconstruction-feasible pose pairs}

\begin{figure*}[t]
  \centering
  \includegraphics[width=0.98\textwidth]{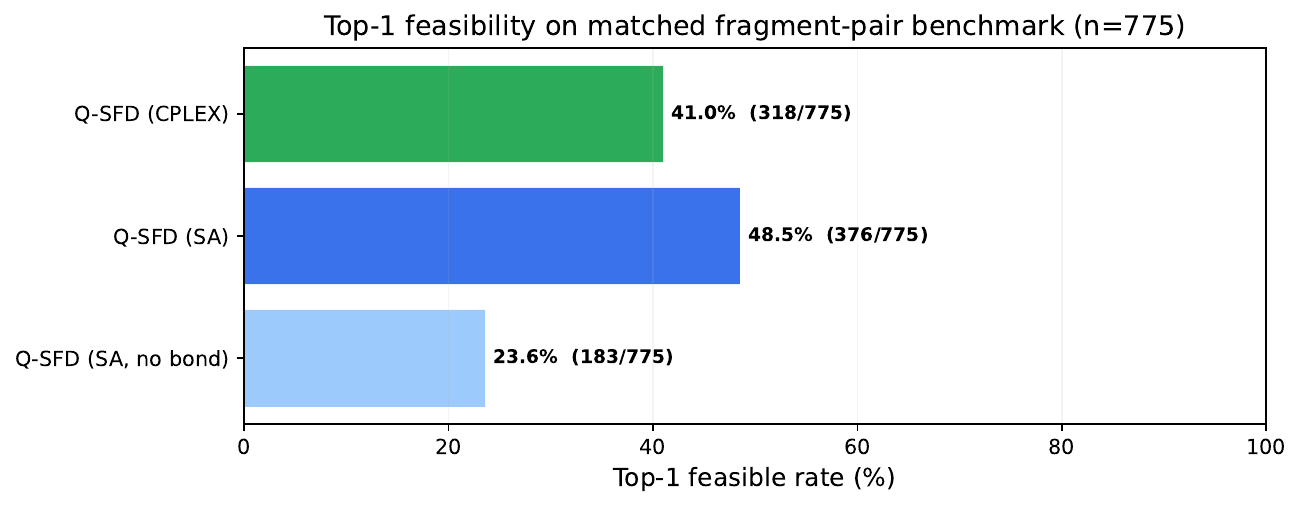}
  \caption{Top-1 recovery of reconstruction-feasible pose pairs on the matched fragment-pair benchmark ($n=775$). Q-SFD~(SA) outperforms the version without the inter-fragment distance term, while exact optimization of the same QUBO by CPLEX shows that objective optimality alone does not fully determine downstream feasibility.}
  \label{fig:top1_feasibility}
\end{figure*}

Top-1 feasibility was compared among Q-SFD solved with simulated annealing (denoted Q-SFD~(SA)), the corresponding version without the inter-fragment distance term, and CPLEX across all 775 benchmark cases (Figure~\ref{fig:top1_feasibility}). Q-SFD~(SA) achieved a top-1 feasibility of 48.5\% (376/775), whereas the version without the distance term reached 23.6\% (183/775). CPLEX returned a feasible top-1 solution in 41.0\% of cases (318/775). Adding the inter-fragment distance term therefore approximately doubled top-1 recovery relative to the ablated formulation.

The lower top-1 feasibility of CPLEX does not indicate a failure of exact optimization. CPLEX minimizes the QUBO objective exactly, whereas feasibility is evaluated after full-atom reconstruction using separate geometric criteria. The exact QUBO optimum therefore need not coincide with the reconstruction-feasible optimum under the downstream criterion. In this setting, simulated annealing can return a different low-energy state whose reconstructed geometry better satisfies that downstream criterion, even when that state is not the exact QUBO minimum. Additional solver-role analyses are provided in Appendix~\ref{app:si}.

The practical effect of retaining multiple low-energy states becomes clearer at larger $K$, where $K$ denotes the number of retained solutions. The fraction of cases with at least one feasible solution rose from 48.5\% at top-1 to 92.6\% at $K\!=\!5$ (Fig.~\ref{fig:s1_rankk}). Because CPLEX returns a single deterministic optimum, its feasibility does not change with $K$.

For context, AutoDock Vina runs under the same shared grid box are summarized in Table~\ref{tab:contextual_baselines}~\cite{eberhardt2021_vina120,Trott2010Vina}. Individual-fragment docking yielded 1.9\% (15/775), whereas simultaneous multi-ligand docking yielded 0/775. 

\subsection{The inter-fragment distance term reshapes the solution distribution toward reconstruction-feasible geometry}

\begin{figure*}[!t]
  \centering
  \includegraphics[width=\textwidth]{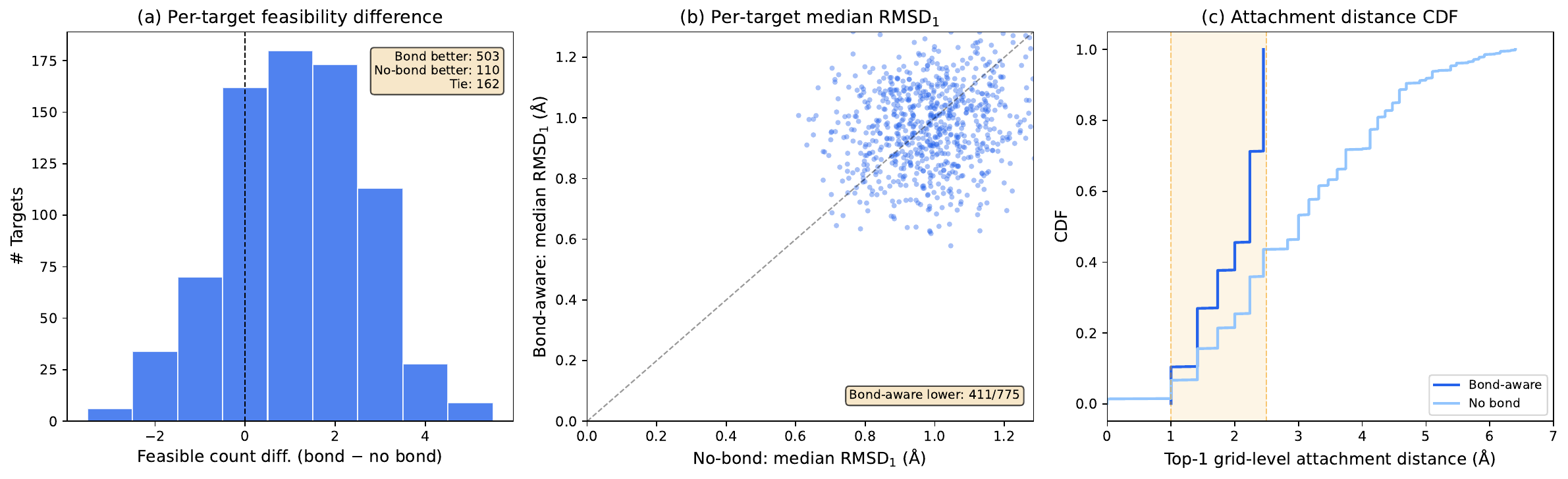}
  \caption{Mechanistic effect of the inter-fragment distance term. (a) Per-case difference in the number of feasible reconstructed solutions in the retained solution pool between Q-SFD~(SA) and the version without the distance term; positive values favor the distance-constrained formulation. (b) Per-case comparison of median RMSD$_1$ across retained reconstructed solutions; points below the diagonal favor the distance-constrained formulation. (c) Empirical cumulative distribution of the top-1 reconstructed attachment distance, showing that the distance-constrained formulation shifts returned solutions toward shorter, bond-compatible geometry.}
  \label{fig:distance_ablation}
\end{figure*}

Two mechanisms could explain the feasibility improvement. The inter-fragment distance term could act only as a downstream filter that rejects non-linkable outcomes after optimization, or it could reshape the optimization landscape so that the solver returns qualitatively different placements. Per-case analysis supported the latter interpretation (Figure~\ref{fig:distance_ablation}).

Across the 775 benchmark cases, the distance-constrained formulation recovered more feasible reconstructed solutions in the retained pool than the version without the distance term in 503 cases (the ablated version was superior in 110; tied in 162). The distance-constrained formulation also produced a lower per-case median RMSD$_1$ in 411 cases, compared with 364 cases for the ablated version. These trends indicate that the inter-fragment distance term actively steers the returned solution set toward attachment-compatible geometry rather than simply rejecting poor solutions after optimization.

The top-1 reconstructed attachment-distance distribution supports the same conclusion. Under the distance-constrained formulation, top-1 solutions were concentrated within or near the benchmark bond interval, whereas the version without the distance term exhibited a broader tail toward longer reconstructed attachment distances.

\subsection{Fragment pose accuracy is preserved while feasibility improves}

\begin{figure*}[!t]
  \centering
  \includegraphics[width=\textwidth]{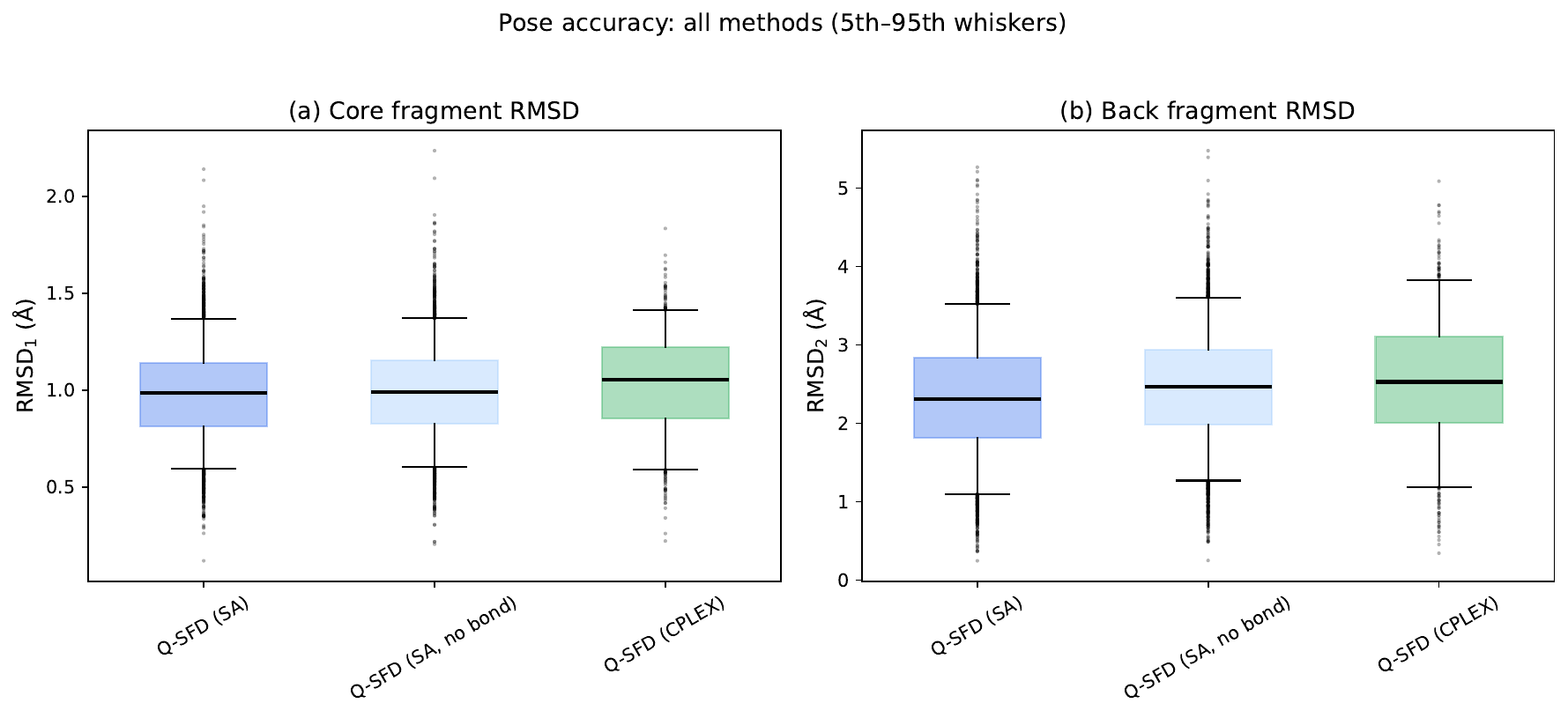}
  \caption{Reconstructed pose accuracy on the matched fragment-pair benchmark ($n=775$). (a) RMSD$_1$ distributions remain tightly clustered across all three conditions, with median values close to 1\,\AA. (b) RMSD$_2$ distributions are broader, consistent with the larger candidate budget assigned to fragment~2, but the distance-constrained formulation does not degrade fragment-level pose fidelity relative to the version without the distance term.}
  \label{fig:rmsd}
\end{figure*}

We next examined whether the feasibility improvement came at the expense of fragment-level pose fidelity (Figure~\ref{fig:rmsd}). For fragment~1, Q-SFD~(SA) achieved a median root-mean-square deviation (RMSD$_1$) of 0.97\,\AA, compared with 0.99\,\AA\ for the version without the distance term. For fragment~2, the corresponding median RMSD$_2$ values were 2.31\,\AA\ and 2.46\,\AA. Adding the inter-fragment distance term therefore improved feasibility without degrading fragment-level pose accuracy.

The broader RMSD$_2$ distribution is consistent with the larger candidate budget allocated to fragment~2. Because this fragment explores a wider spatial region, its reconstructed pose is inherently more variable. Crucially, the distance-constrained formulation did not increase RMSD$_2$ relative to the ablated version, confirming that the feasibility gain does not come at the cost of pose fidelity. CPLEX is included in Figure~\ref{fig:rmsd} as an objective-optimal reference for the same QUBO.

\subsection{A supplementary hybrid follow-up recovers additional feasible solutions in a stringent subset}

\begin{figure*}[!t]
  \centering
  \includegraphics[width=0.98\textwidth]{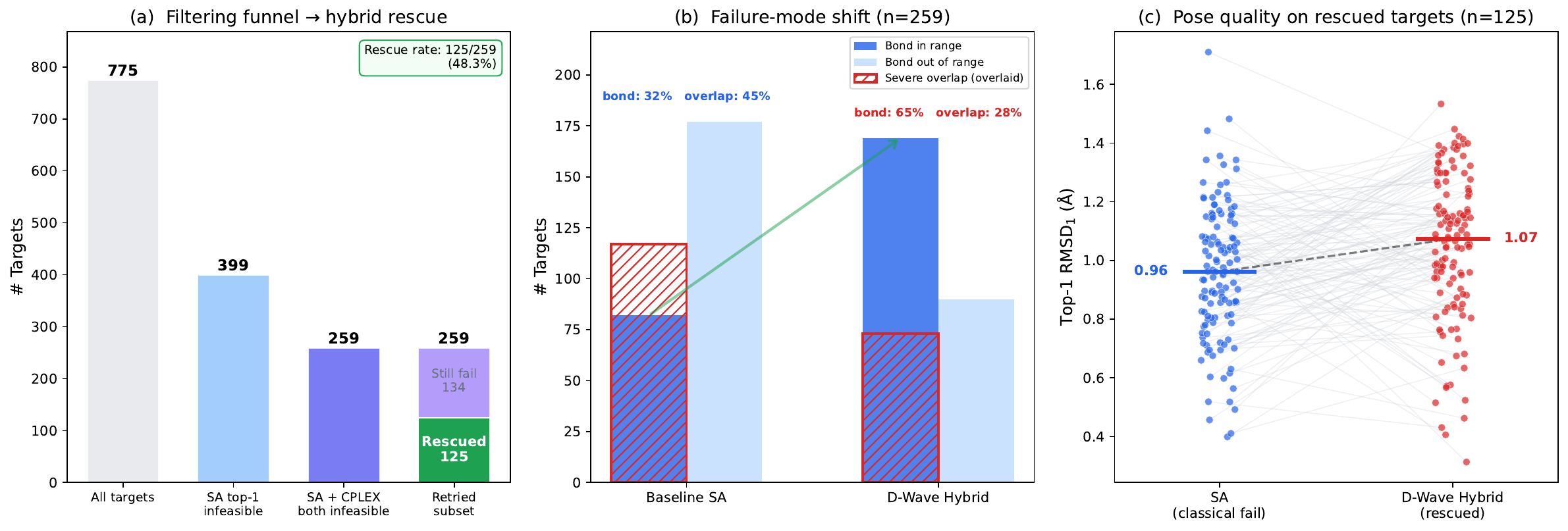}
  \caption{Supplementary hybrid follow-up with the D-Wave Leap hybrid solver on the 259 cases that remained top-1 infeasible under both baseline Q-SFD~(SA) and CPLEX. (a) Rescue count within this stringent subset. (b) Shift in predicted attachment distance from baseline simulated annealing to hybrid optimization. (c) Top-1 RMSD$_1$ comparison for the rescued cases. (d) Summary of failure modes for the same subset; severe overlap is overlaid and is not mutually exclusive with bond-range status.}
  \label{fig:hybrid_retry}
\end{figure*}

We performed a supplementary follow-up using the D-Wave Leap hybrid solver on the 259 cases that remained top-1 infeasible under both Q-SFD~(SA) and CPLEX (Figure~\ref{fig:hybrid_retry}). This analysis was not part of the matched main benchmark. It was used to test whether an alternative backend could recover feasible solutions from a deliberately difficult subset under the same formulation. Among these cases, the hybrid solver recovered a feasible top-1 solution in 125 cases (48.3\%).

On the rescued cases, the median RMSD$_1$ shifted from 0.96\,\AA\ for baseline Q-SFD~(SA) to 1.07\,\AA\ for the hybrid solutions, indicating that the additional recovery did not require a substantial loss of fragment fidelity. Figure~\ref{fig:hybrid_retry}d further shows that the hybrid follow-up changed the distribution of failure modes: the fraction of top-1 solutions with attachment distance in range increased, whereas the fraction with severe overlap decreased.

We interpret this result as a supplementary rescue analysis rather than as the main benchmark claim. The main conclusion of the study remains that the inter-fragment distance term improves recovery of reconstruction-feasible pose pairs. The hybrid follow-up shows that, under the same formulation, an alternative backend can recover additional feasible solutions from a difficult subset of cases. The rationale for this analysis and related backend-specific settings is described in Appendix~\ref{app:si}.

\subsection{Kinase case study: domain priors enable practical linkable seed generation}

\begin{figure*}[!t]
  \centering
  \includegraphics[width=0.98\textwidth]{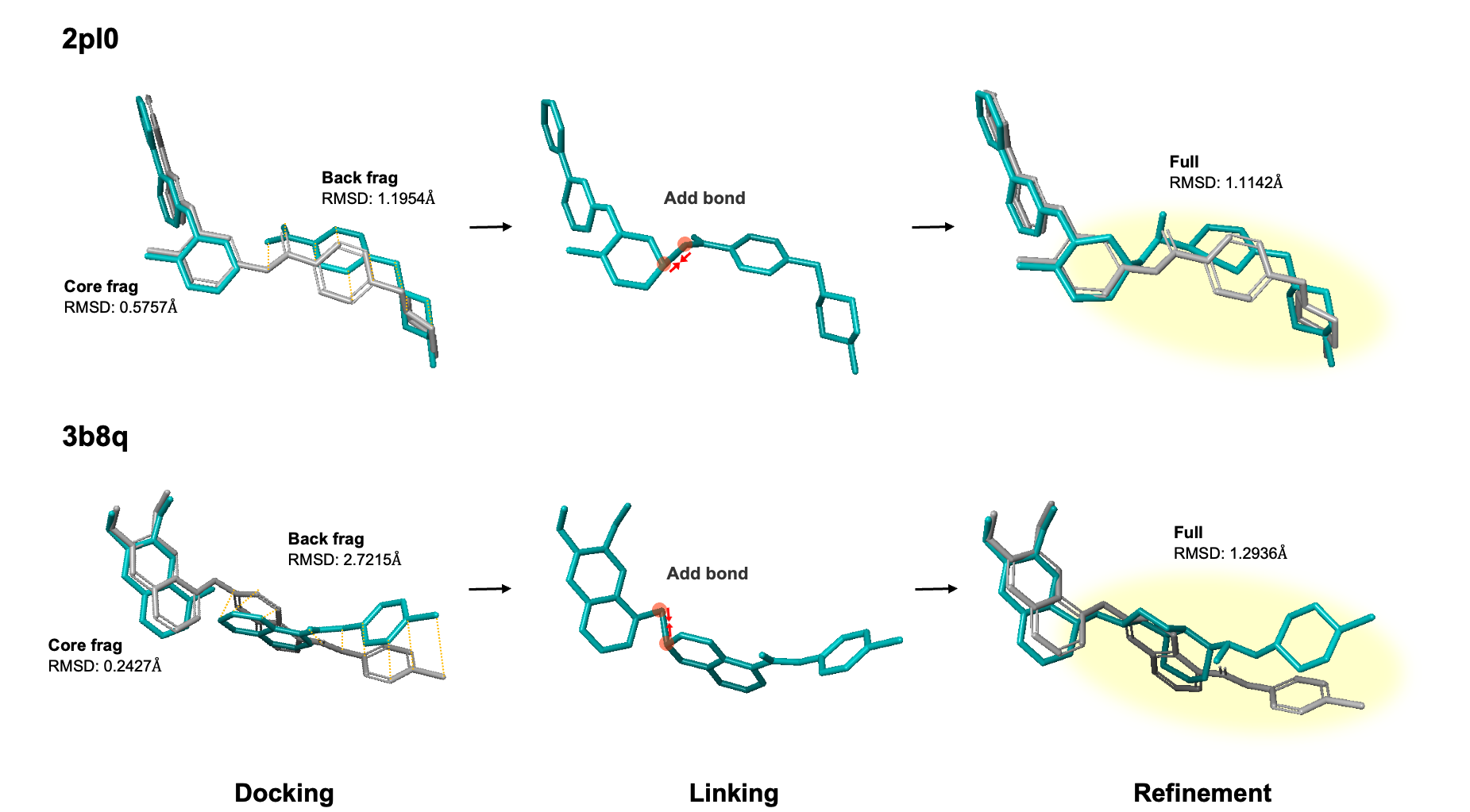}
  \caption{Deployment workflow for the kinase case study on representative Asp-Phe-Gly-out (DFG-out) complexes (PDB~2PL0 and 3B8Q). The panel illustrates how domain priors reshape the search by restricting CORE to a hinge-consistent neighborhood while allocating a broader BACK exploration region, followed by docking, auxiliary linking, and downstream refinement. In this illustrative case study, the docking-stage QUBOs were optimized with the D-Wave Leap hybrid solver. For 2PL0, the CORE, BACK, and refined full-ligand RMSDs were 0.58, 1.20, and 1.11\,\AA, respectively. For 3B8Q, the corresponding values were 0.24, 2.72, and 1.29\,\AA.}
  \label{fig:kinase_case}
\end{figure*}

To illustrate domain-guided deployment under the same formulation, we applied a hinge-guided protocol using a hinge-binding fragment (CORE) and a back-pocket-extending fragment (BACK) to two representative Asp-Phe-Gly-out (DFG-out) kinase complexes, PDB~2PL0 and 3B8Q (Figure~\ref{fig:kinase_case}). In this illustrative analysis, the docking-stage QUBOs were optimized with the D-Wave Leap hybrid solver. CORE candidates were restricted to a hinge-consistent neighborhood, while BACK candidates were given a broader exploration budget toward the expanded pocket characteristic of the DFG-out state. Constraint penalties were adjusted on the basis of target-specific interaction patterns.

Figure~\ref{fig:kinase_case} summarizes the workflow in three stages. In the \emph{Docking} step, CORE and BACK placements are obtained with Q-SFD, and fragment-level RMSDs are reported. In the \emph{Linking} step, an auxiliary bond is introduced between the two attachment atoms to preserve the docked relative placement during downstream geometry processing; this bond is a temporary construct for refinement and evaluation, not a chemically finalized linkage. In the \emph{Refinement} step, the connected intermediate is relaxed and then aligned to the crystallographic full ligand for RMSD evaluation.

For 2PL0, the CORE, BACK, and refined full-ligand RMSDs were 0.58, 1.20, and 1.11\,\AA, respectively. For 3B8Q, the corresponding values were 0.24, 2.72, and 1.29\,\AA. The larger BACK RMSD for 3B8Q reflects the broader exploration region allocated to that fragment, yet the refined full ligand still remained close to the crystallographic reference. These results illustrate that, when domain priors are used to shape the search space, Q-SFD can produce practically useful linkable seeds for downstream medicinal-chemistry workflows.

\section{Methods}

\subsection{Problem definition and benchmark scope}

The task studied here is not generic simultaneous placement of multiple ligands, but simultaneous placement of two rigid fragments in a shared binding site such that they remain geometrically compatible with later covalent connection into a single ligand. The required output is therefore not a pair of individually plausible fragment poses, but a \emph{reconstruction-feasible pose pair}.

Because each benchmark case is generated by retrospective cleavage of a crystallographic ligand, the present study evaluates a matched fragment-pair reconstruction problem rather than \emph{de novo} fragment linking from independently observed fragment hits. Local candidate neighborhoods are constructed around the reference fragment geometry, providing a controlled test of whether the optimization objective can recover linkable two-fragment geometry under known reference binding arrangements.

\subsection{Dataset and fragment generation}

Protein--ligand complexes were collected from the CASF-2016 benchmark set~\cite{su2019casf}, which is based on structures available from the Protein Data Bank~\cite{Berman2000PDB}. Entries were restricted to X-ray diffraction structures with resolution $\leq 2.5$\,\AA. Receptors were constructed from ATOM records only; crystallographic waters, ions, and buffer molecules were removed. Receptor and ligand inputs were converted into AutoDock-compatible formats for grid-map generation and QUBO construction~\cite{Morris2009AutoDock4,OBoyle2011OpenBabel}.

Each ligand was decomposed into two rigid fragments by cutting a single non-ring rotatable bond. Candidate cut bonds were restricted to non-ring single bonds whose endpoints are non-terminal heavy atoms, and only two-fragment outcomes were retained. Both fragments were required to satisfy a heavy-atom count range of $3 \leq N_{\mathrm{heavy}} \leq 10$; among feasible cuts, the most size-balanced decomposition was selected. The heavy atoms at the two ends of the cleaved bond were designated as the attachment atom $a^{*}$ of fragment~1 and the attachment atom $b^{*}$ of fragment~2, respectively. Cases were excluded when fragmentation, AutoGrid4 map generation, or detection of the cleaved linking bond failed during preprocessing.

Of the 2,647 complexes in CASF-2016, 775 passed all preprocessing steps and were included in the final benchmark. Each of these 775 cases corresponds to a unique PDB structure.

\subsection{Shared-grid setup and candidate construction}

Both fragments were evaluated on a single shared docking grid. The grid center was set at the midpoint of the two attachment-atom coordinates; the grid dimensions were $50.0 \times 50.0 \times 50.0$\,\AA\ with a spacing of 1.0\,\AA. AutoGrid4 was used to precompute van der Waals interaction maps, an electrostatic map, and a desolvation map~\cite{Goodford1985,Morris2009AutoDock4}.

Q-SFD uses an anchor-based representation to keep the QUBO lightweight while preserving rigid geometry. In the benchmark implementation, each fragment was represented by four anchors, including the designated attachment atom on each side. Candidate construction was asymmetric. Fragment~1 anchors were assigned smaller local candidate sets (up to 20 candidates within a 2.0\,\AA\ neighborhood), whereas fragment~2 anchors were given a broader search budget (up to 20 candidates within a 4.0\,\AA\ neighborhood, and up to 30 candidates for the attachment anchor $b^{*}$). This design keeps one fragment relatively localized while granting the other greater spatial freedom within the same distance-constrained optimization framework.

Because the benchmark is retrospective, these local candidate neighborhoods are constructed around known fragment placements. They are used to test whether the formulation can recover linkable two-fragment geometry under controlled conditions. No RMSD-based filtering or reference-based solution ranking was applied.

\subsection{Q-SFD QUBO formulation}

Q-SFD formulates the simultaneous placement of two rigid fragments as a single quadratic unconstrained binary optimization (QUBO) problem. Related QUBO-based docking formulations have also been reported for fragment-based flexible docking and annealer-oriented molecular docking~\cite{Yanagisawa2024Entropy,Triuzzi2025QST}. Let $\mathcal{G}_a$ denote the set of candidate grid points for anchor $a$, let $r_g$ denote the Cartesian coordinate of candidate point $g$, and let $\mathbb{1}[\cdot]$ denote the indicator function. A binary variable $x_{a,g} \in \{0,1\}$ encodes whether anchor $a$ is placed at candidate grid point $g \in \mathcal{G}_a$.

The QUBO objective is
\begin{equation}\label{eq:qubo}
  \min_{x \in \{0,1\}}
  \underbrace{\sum_{a}\sum_{g \in \mathcal{G}_a} E_{a,g}\,x_{a,g}}_{H_{\mathrm{unary}}}
  \;+
  \underbrace{\lambda_{\mathrm{one}}\sum_{a}\!\Bigl(1 - \sum_{g \in \mathcal{G}_a} x_{a,g}\Bigr)^{\!2}}_{H_{\mathrm{OH}}}
  \;+
  \underbrace{\lambda_{1}\,\Phi_{1}(x) + \lambda_{2}\,\Phi_{2}(x)}_{H_{\mathrm{rigid}}}
  \;+
  \underbrace{\lambda_{\mathrm{dist}}\,\Psi_{\mathrm{dist}}(x)}_{H_{\mathrm{dist}}}.
\end{equation}

\paragraph{Unary interaction term.}
$E_{a,g}$ is the protein--fragment interaction energy read from the AutoGrid4 map corresponding to the atom type of anchor $a$, combining van der Waals, electrostatic ($w_{\mathrm{elec}}=2.0$), and desolvation ($w_{\mathrm{dsolv}}=1.0$) contributions.

\paragraph{One-hot assignment.}
Each anchor must be placed at exactly one candidate ($\lambda_{\mathrm{one}} = 200{,}000$).

\paragraph{Intra-fragment rigid-body constraint.}
Let $\mathcal{P}_f$ denote the set of anchor pairs belonging to fragment $f \in \{1,2\}$. For each anchor pair $(a,a') \in \mathcal{P}_f$, a penalty is incurred when the pairwise distance on the grid deviates from the reference distance $d^{0}_{aa'}$ beyond a tolerance $\mathcal{T}_f$:
\begin{equation}\label{eq:rigid}
  \Phi_f(x) = \!\!\sum_{(a,a') \in \mathcal{P}_f}\;\sum_{g \in \mathcal{G}_a}\;\sum_{g' \in \mathcal{G}_{a'}}
  \mathbb{1}\!\bigl[\,\lvert\|r_g - r_{g'}\| - d^{0}_{aa'}\rvert > \mathcal{T}_f\,\bigr]\;
  x_{a,g}\,x_{a',g'}.
\end{equation}
Fragment~1 uses $\lambda_{1} = 200$ and $\mathcal{T}_1 = 1.0$\,\AA. Fragment~2 uses $\lambda_{2} = 800$ and $\mathcal{T}_2 = 1.5$\,\AA.

\paragraph{Inter-fragment distance term.}
A penalty is applied when the distance between the attachment anchors $a^{*}$ and $b^{*}$ falls outside a fixed chemically plausible interval:
\begin{equation}\label{eq:dist}
  \Psi_{\mathrm{dist}}(x) = \sum_{g \in \mathcal{G}_{a^*}}\;\sum_{g' \in \mathcal{G}_{b^*}}
  \mathbb{1}\!\bigl[\,\|r_g - r_{g'}\| \notin [r_{\min}, r_{\max}]\,\bigr]\;
  x_{a^*\!,g}\,x_{b^*\!,g'}.
\end{equation}
In the main benchmark, we use $[r_{\min}, r_{\max}] = [1.0, 2.5]$\,\AA\ and $\lambda_{\mathrm{dist}} = 40{,}000$. The same interval is used in evaluation. This term is the defining contribution of Q-SFD and is validated by comparison with an ablation that sets $\lambda_{\mathrm{dist}}=0$.

Because the QUBO is a quadratic polynomial in binary variables, the same objective can be solved by classical heuristics, exact solvers, and hybrid quantum--classical workflows~\cite{Lucas2014,glover2019_qubo,Kochenberger2014,mcgeoch2014aqc,aramon2019_digitalannealer}. This solver-agnostic property is one of the primary motivations for casting Q-SFD as a QUBO.

\subsection{Reconstruction and evaluation metrics}

Binary solutions were decoded into anchor assignments, and a Kabsch rigid-body alignment~\cite{Kabsch1976} was applied to each fragment to reconstruct full-atom coordinates. Success was judged not by docking score or QUBO energy, but by post-reconstruction geometric criteria. A solution was classified as feasible if it simultaneously satisfied two conditions: (i) the reconstructed attachment distance $d_{\mathrm{link}} \in [1.0, 2.5]$\,\AA\ (\emph{linkability}), and (ii) the minimum inter-fragment heavy-atom distance $d_{\min} \geq 0.8$\,\AA\ (\emph{no severe overlap}). Throughout the manuscript, feasibility denotes the joint condition of attachment distance in range and no severe overlap.

Fragment-level RMSD values (RMSD$_1$, RMSD$_2$) were computed against the crystallographic reference. We additionally report per-case differences in feasible-solution counts and per-case median RMSD comparisons to characterize how the inter-fragment distance term changes the solution distribution. Before reconstruction, one-hot validity and anchor completeness were explicitly verified so that invalid decoded states were not interpreted as geometric solutions.

\subsection{Solvers, baselines, and resource logging}

\paragraph{Primary solver (simulated annealing).}
The main Q-SFD benchmark used classical simulated annealing (SA)~\cite{Kirkpatrick1983} with multiple reads; the top-5 solutions were retained for each case. Retaining a pool of near-optimal states allows downstream reconstruction to exploit solution diversity, which is especially useful when objective optimality does not uniquely determine geometric feasibility.

\paragraph{Exact reference (IBM CPLEX).}
To separate formulation effects from solver effects, we additionally solved the same QUBO with IBM CPLEX as an exact reference for the top-1 objective minimum. Because reconstruction feasibility is evaluated outside the QUBO objective, the exact solver is used here to interpret how the current formulation relates to the downstream criterion rather than as a presumed upper bound on downstream reconstruction feasibility.

\paragraph{Contextual docking references (AutoDock Vina).}
As contextual references, we evaluated AutoDock Vina using the same shared grid box as Q-SFD, including both individual docking of the two fragments and simultaneous multi-ligand docking~\cite{eberhardt2021_vina120,Trott2010Vina}. We also performed a separate whole-ligand docking run on the larger intact-ligand set available before retrospective fragmentation. These runs were not intended as task-matched primary baselines. They are reported only to show how related docking tasks behave under a conventional docking tool that optimizes docking score rather than post-reconstruction linkability.

\paragraph{Supplementary hybrid follow-up (D-Wave Leap).}
As a supplementary analysis, we applied the D-Wave Leap hybrid solver to the 259 cases that were top-1 infeasible under both SA and CPLEX. This follow-up was used as a rescue analysis under the same formulation rather than as a population-level benchmark against the main 775-target SA/CPLEX comparison. The rationale for this hybrid run, together with backend-specific settings and additional solver-role analyses, is described in Appendix~\ref{app:si}~\cite{Stogiannos2022Hybrid,Hauke2020,Yarkoni2022}.

\paragraph{Resource logging.}
Resource usage was logged at two levels. The main benchmark records logical instance quantities such as variable count, shared-grid construction time, QUBO build time, and solve time. The supplementary hybrid follow-up reports rescue-analysis results on the stringent classical-failure subset. For hybrid runs, hardware-level quantities such as physical-qubit count, chain length, and quantum processing unit (QPU) access time are not uniformly exposed by the solver and are therefore reported only when available.

\subsection{Kinase case study protocol}

In structure-based drug design, domain priors about the target are routinely used to reshape the search space. To illustrate how Q-SFD can be deployed in such a setting, we conducted a case study on two representative Asp-Phe-Gly-out (DFG-out) kinase complexes, PDB~2PL0 and 3B8Q.

Each ligand was decomposed into a hinge-binding fragment (CORE) and a back-pocket-extending fragment (BACK). CORE candidates were restricted to a hinge-consistent local neighborhood, while BACK was allocated a larger exploration budget toward the expanded hydrophobic region characteristic of the DFG-out conformation~\cite{Xing2015,ZhaoBourne2023,Vijayan2015,Modi2019,LiuGray2006,Kufareva2008,Attwood2021,Cohen2021}. Candidates for the BACK attachment anchor $b^{*}$ were generated from a spherical-shell region around $a^{*}$ with radii $[r_{\min}, r_{\max}]$, so that bond-length feasibility is reflected directly in the candidate pool. A CORE-exclusion rule removed BACK candidates within a small radius of CORE atoms to suppress unrealistic placements.

The QUBO instances were optimized with the D-Wave Leap hybrid solver rather than the SA/CPLEX benchmark workflow, because the purpose was to illustrate domain-guided deployment under limited candidate and QUBO budgets rather than to perform matched solver benchmarking. Constraint penalties were adjusted on the basis of target-specific interaction patterns. After docking, a temporary single bond was added between the two attachment atoms to preserve the recovered relative placement during downstream geometry processing. The connected intermediate was then relaxed, and the refined full ligand was aligned to the crystallographic ligand to compute the reported full-ligand RMSD. This case study is not pooled with the main benchmark statistics.

\subsection{Supplementary solver-backend analyses}

Additional analyses are placed in Appendix~\ref{app:si} to clarify calibration sensitivity, target dependence, and solver roles under limited quantum-computing resources. These include an earlier shared-grid simulated annealing--quantum processing unit (SA--QPU) benchmark on a smaller set, a QPU penalty-scale sweep, and a hybrid-versus-QPU reference comparison.

\section{Discussion}

In this study, we addressed the problem of recovering fragment pose pairs that remain compatible with later covalent connection. Rather than treating fragment placement and linkability assessment as separate steps, we formulated simultaneous placement of two rigid fragments as a single QUBO problem and encoded geometric linkability directly in the objective through an explicit inter-fragment distance term. This formulation was designed to favor pose pairs that are not only individually favorable within the binding site but also suitable for reconstruction into a single molecule.

This design choice is important because fragment linking requires more than identifying plausible poses for the individual fragments. A useful computational solution must preserve both local protein compatibility and inter-fragment geometry after reconstruction. Conventional fragment-docking workflows can generate reasonable poses for individual fragments, but they do not directly optimize this linking-oriented placement problem~\cite{Kitchen2004,Pagadala2017,Warren2006}. In that sense, the present study is best understood as a focused formulation study for simultaneous fragment placement under an explicit geometric linkability criterion.

The solver comparison provides a complementary view of this mechanism. Per-case analyses showed that the inter-fragment distance term reshaped the returned solution distribution rather than acting merely as a downstream filter. The attachment-distance distribution shifted toward shorter, bond-compatible values. At the same time, CPLEX, which exactly minimizes the same QUBO objective, achieved lower top-1 reconstruction feasibility than simulated annealing. This pattern reflects the fact that objective optimality and downstream geometric feasibility are related but distinct criteria. The exact QUBO minimum does not necessarily correspond to the reconstruction-feasible optimum after full-atom rebuilding. In this setting, simulated annealing can return a different low-energy state whose reconstructed geometry better satisfies the downstream criterion. When multiple low-energy states are retained, this practical advantage becomes more apparent at larger $K$. For fragment-linking applications, solution diversity is therefore a useful property that complements objective quality.

The supplementary hybrid analysis and the kinase case study extend this picture without changing the main conclusion. The hybrid follow-up showed that, under the same distance-constrained formulation, an alternative backend could recover additional feasible solutions from cases that remained difficult under both simulated annealing and CPLEX. The kinase case study showed that target-specific domain priors can be used to reshape the search space and generate practically useful linkable seeds in a medicinal-chemistry setting.

The present study also highlights the value of the QUBO representation itself. Related work has shown that QUBO formulations can also support fragment-based flexible docking and annealer-oriented molecular docking in broader settings~\cite{Yanagisawa2024Entropy,Triuzzi2025QST}. In the present study, the problem is expressed in a form that can be examined with multiple classical and hybrid backends without reformulating the underlying objective~\cite{Lucas2014,glover2019_qubo,Hauke2020,Yarkoni2022}. That flexibility is useful even at the present benchmark scale. It may become more important as fragment-placement problems grow in size and complexity, because the same formulation can be transferred across solver backends while preserving the same physical interpretation.

The study has clear limitations. The benchmark is retrospective, and each case was generated by cleaving a crystallographic ligand. The present results therefore test whether the method can recover the correct fragment geometry under controlled conditions. All fragments were treated as rigid bodies. Inter-fragment steric compatibility was still evaluated after reconstruction rather than fully encoded inside the QUBO. These boundaries mean that the current work is best interpreted as a focused formulation study on simultaneous fragment placement.

These limitations define the most immediate next steps. A natural extension is prospective evaluation on independently observed fragment-hit pairs. Encoding steric compatibility more directly in the objective would bring the QUBO target into closer alignment with downstream feasibility. End-to-end integration with linker-generation workflows would make it possible to assess how much this upstream placement stage improves fragment-to-lead performance in practice.

\section{Conclusion}

In this study, we presented Q-SFD, a QUBO-based framework for linking-oriented fragment placement. Rather than assessing geometric linkability as a downstream filter, Q-SFD embeds it directly in the optimization objective through an explicit inter-fragment distance term, so that the returned pose pairs favor both local protein compatibility and mutual covalent reconstructability from the outset.

On a retrospective matched fragment-pair benchmark, the distance-constrained formulation approximately doubled top-1 recovery of reconstruction-feasible pose pairs relative to the version without this term. Considering the top-5 solutions increased recovery to more than 90\% of benchmark cases. These gains were achieved without degrading fragment-level pose accuracy. Together, these results show that directly optimizing linkability is an effective strategy for linking-oriented fragment placement.

Q-SFD is intended for the upstream stage of fragment-linking workflows. It provides geometrically linkable fragment-pair seeds before linker design begins. In that role, it offers both a practical implementation and a clear formulation principle: geometric linkability is most effective when it is treated as part of the optimization target itself.

\bmhead{Acknowledgements}

This work received partial support from the following sources:
(1) Basic Science Research Program through the National Research Foundation of Korea (NRF), funded by the Ministry of Science and ICT (RS-2023-NR068116, RS-2025-03532992).
(2) Institute for Information \& Communications Technology Promotion (IITP) grant funded by the Korea government (MSIP) (No.\ 2019-0-00003), which focuses on the research and development of core technologies for programming, running, implementing, and validating fault-tolerant quantum computing systems.
(3) Korean ARPA-H Project via the Korea Health Industry Development Institute (KHIDI); Ministry of Health and Welfare, Republic of Korea (RS-2025-25456722).
(4) Yonsei University Research Fund under project number 2025-22-0140.

\bmhead{Declarations}

\begin{itemize}
\item \textbf{Funding} National Research Foundation of Korea, Institute for Information \& Communications Technology Promotion, Korea Health Industry Development Institute, Yonsei University.
\item \textbf{Author contribution}
J.~L. performed the experiments and drafted the manuscript.
Y.~K.~C. reviewed and provided feedback on the manuscript.
J.~H. conceived the idea, supervised the project, and reviewed the manuscript.
\end{itemize}

\bigskip
\clearpage

\begin{appendices}

\section{Supporting Information}\label{app:si}

\setcounter{figure}{0}
\renewcommand{\thefigure}{S\arabic{figure}}
\setcounter{table}{0}
\renewcommand{\thetable}{S\arabic{table}}

\subsection{Overview and scope}\label{app:overview}

This Supporting Information provides additional methodological detail and supplementary analyses supporting the main text. Throughout this document, Q-SFD denotes QUBO-based simultaneous fragment docking. Unless otherwise stated, feasibility is evaluated after full-atom reconstruction using two criteria simultaneously: (i)~the reconstructed attachment distance must lie within the fixed benchmark bond interval of 1.0--2.5\,\AA, and (ii)~the reconstructed fragments must not exhibit severe heavy-atom overlap. Throughout this document, the phrase \emph{inter-fragment distance term} refers to the QUBO term that penalizes attachment-anchor distances outside this interval.

The main benchmark in the manuscript was designed to test the effect of the QUBO formulation under a single matched protocol across all 775 targets. The additional backend analyses reported here were performed on different subsets or under backend-specific settings. They are therefore provided to clarify solver behavior under the same general formulation, rather than to redefine the main 775-target benchmark.

\subsection{Shared-grid protocol and reconstruction-based evaluation}\label{app:shared_grid}

All solvers in the shared-grid benchmark use the same docking box and the same AutoGrid-derived interaction maps, so differences between methods arise from optimization behavior rather than from differences in scoring inputs. After decoding a binary solution, each fragment is reconstructed by rigid alignment from the selected anchor placements to the corresponding full-atom fragment coordinates. This reconstruction step is essential because anchor-level satisfaction alone does not guarantee that the two reconstructed fragments remain geometrically compatible at the all-atom level.

Throughout this Supporting Information, a solution in the 775-target matched benchmark is considered feasible only when the reconstructed attachment distance satisfies the fixed benchmark bond interval of 1.0--2.5\,\AA\ and the reconstructed minimum inter-fragment heavy-atom distance is at least 0.8\,\AA. The same criteria are used in the main text to interpret feasibility, attachment-distance behavior, and reconstructed fragment RMSDs.

\subsection{Decoded-state validity checks}\label{app:validity}

Before full-atom reconstruction is performed, each decoded binary solution is verified against two conditions. First, one-hot validity is checked: for each anchor~$a$, exactly one candidate grid point must be assigned, i.e.\ $\sum_{g \in \mathcal{G}_a} x_{a,g} = 1$. Solutions in which any anchor is unassigned or multiply assigned are marked invalid and excluded from reconstruction. Second, anchor completeness is verified: all anchors required to define the rigid placement of each fragment must be present in the decoded state. A solution is accepted for reconstruction only when both conditions are satisfied for every anchor in both fragments.

These checks are applied uniformly across all solvers and all benchmark targets. Because the one-hot penalty $H_{\mathrm{OH}}$ carries a large coefficient ($\lambda_{\mathrm{one}} = 200{,}000$), one-hot violations are strongly disfavored, although they can still arise in low-quality SA reads or in edge cases where the QUBO landscape is relatively flat. Reporting these checks explicitly ensures that feasibility counts reflect only geometrically valid reconstructions rather than decoding failures.

\subsection{Solver configurations and resource reporting}\label{app:solver_config}

The primary benchmark uses classical simulated annealing (SA) with multiple reads and retention of the top-5 decoded solutions for downstream evaluation. Exact optimization with CPLEX is included as a same-QUBO reference for the top-1 objective minimum. These two solvers define the matched main-benchmark comparison because they allow the effect of the formulation to be examined under a single fixed protocol across all 775 targets.

The supplementary analyses below include an earlier shared-grid SA--QPU benchmark, a QPU penalty-scale sweep, a hybrid-versus-QPU reference comparison, and a hybrid rescue analysis on the stringent classical-failure subset. Resource usage is reported at two levels. Logical instance statistics such as variable count, QUBO build time, and solve time are summarized for the main benchmark. Hardware-level quantities such as physical-qubit count, chain statistics, and QPU access time are reported when available. For D-Wave Leap hybrid runs, these hardware-level quantities are not directly exposed by the solver and are therefore reported as unavailable.

The supplementary hybrid follow-up discussed in the main text was analyzed on the stringent subset of 259 targets that remained top-1 infeasible under both SA and CPLEX. Completed hybrid run files also contained additional SA-only failure cases, but those were excluded from the rescue analysis reported in the manuscript. Accordingly, all hybrid rescue statements and resource summaries in this Supporting Information refer only to the 259-target strict subset.

\subsection{Additional analyses for the 775-target matched benchmark}\label{app:note_s1}

\begin{figure}[!htbp]
  \centering
  \includegraphics[width=0.70\linewidth]{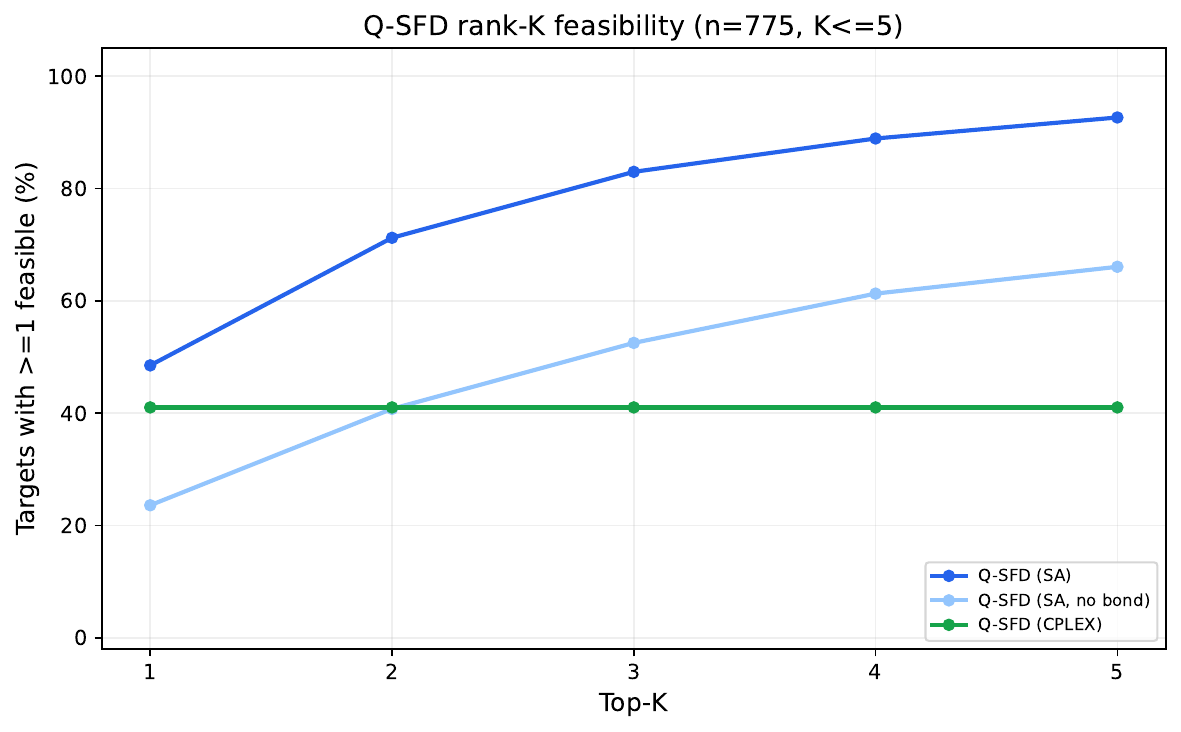}
  \caption{Rank-$K$ feasibility of the Q-SFD variants on the matched fragment-pair benchmark ($n=775$, $K\leq 5$). The steep increase for Q-SFD~(SA) from $K{=}1$ to $K{=}5$ highlights the practical value of retaining multiple near-optimal solutions.}
  \label{fig:s1_rankk}
\end{figure}

Figure~\ref{fig:s1_rankk} reports rank-$K$ feasibility for the 775-target matched fragment-pair benchmark. While the main text emphasizes top-1 behavior, the retained multi-solution pool is also practically important because top-$K$ recovery rises steeply with rank and reaches 92.6\% at $K=5$. This behavior supports the use of multiple retained near-optimal solutions rather than relying exclusively on the top-1 decoded state. In the present setup, CPLEX returns a single deterministic optimum, so its feasibility does not change with rank~$K$.

\FloatBarrier

\begin{figure}[!htbp]
  \centering
  \includegraphics[width=0.78\linewidth]{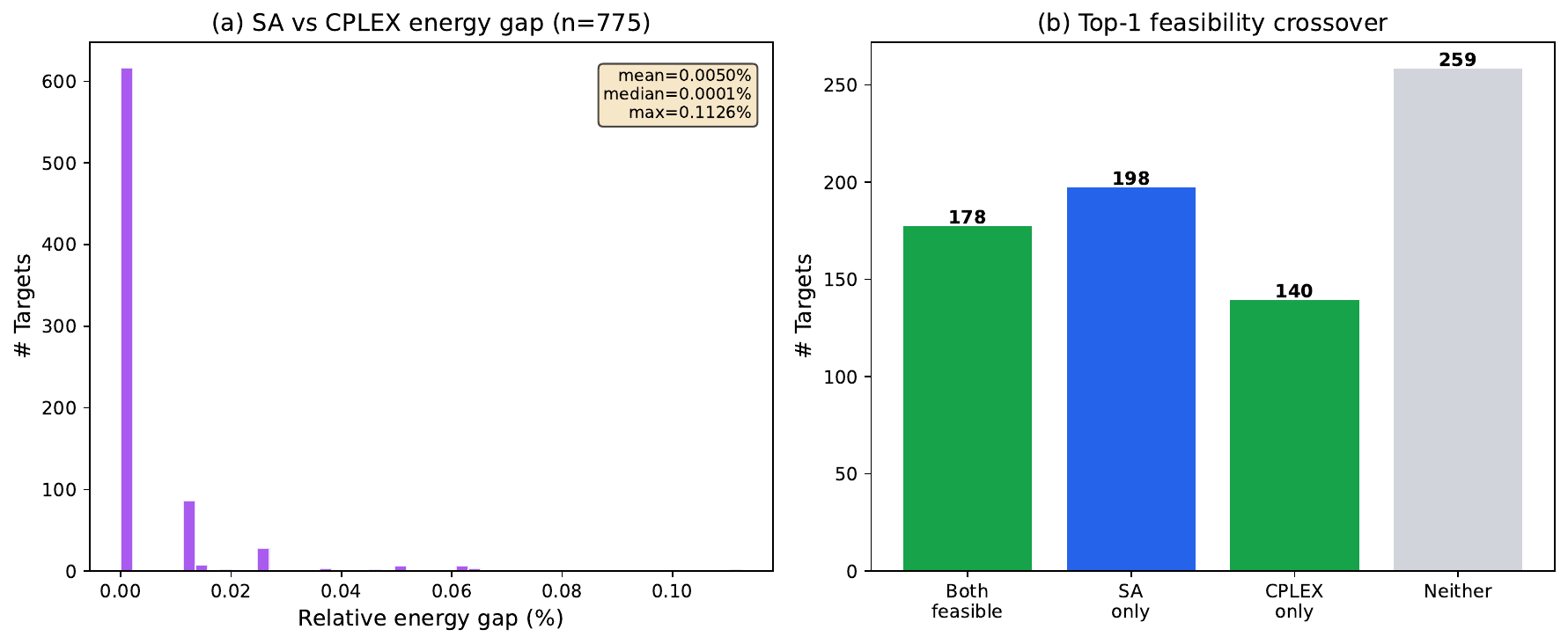}
  \caption{Solver-level comparison between multi-read SA and exact CPLEX on the same QUBO. The two solvers optimize the same formulation, yet downstream reconstruction feasibility can still differ across targets.}
  \label{fig:s2_sa_cplex}
\end{figure}

Figure~\ref{fig:s2_sa_cplex} compares multi-read SA and exact CPLEX on the same QUBO. The two solvers optimize the same formulation but can still differ in downstream reconstruction feasibility across targets.

\FloatBarrier

\subsection{Earlier shared-grid SA--QPU benchmark}\label{app:note_s2}

\begin{figure}[!htbp]
  \centering
  \includegraphics[width=0.80\linewidth]{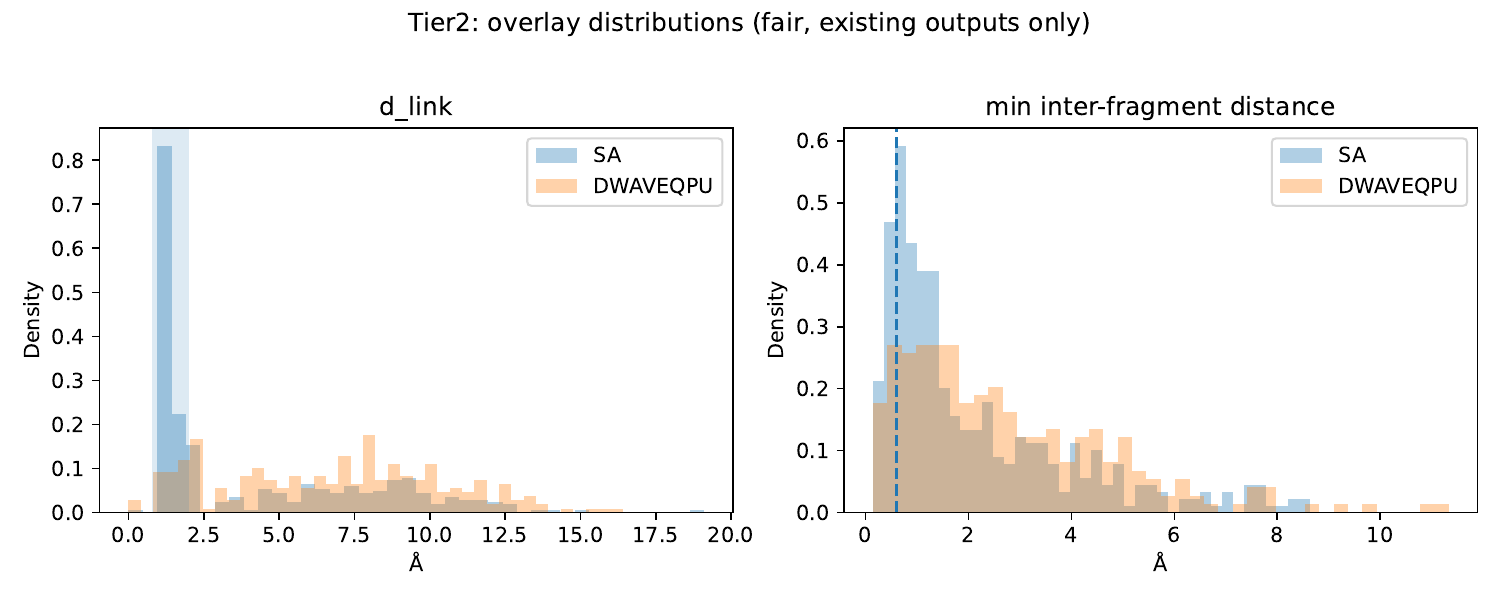}
  \caption{Earlier shared-grid SA--QPU benchmark on a smaller Tier-2 set. Shown are overlay distributions for reconstructed attachment distance ($d_{\mathrm{link}}$) and minimum inter-fragment distance. This analysis reflects a different scale and protocol from the 775-target matched benchmark used in the main text.}
  \label{fig:s3_earlier_qpu}
\end{figure}

Figure~\ref{fig:s3_earlier_qpu} summarizes an earlier shared-grid SA--QPU benchmark carried out on a smaller Tier-2 set. This analysis predates the full 775-target matched benchmark and is included here as supplementary evidence rather than as part of the pooled main benchmark statistics. The figure overlays reconstructed attachment-distance distributions and minimum inter-fragment distance distributions for SA and direct QPU outputs under a fair comparison based on existing outputs only.

Because this earlier SA--QPU analysis was performed on a different benchmark scale and is more sensitive to backend-specific calibration, it is used here to illustrate solver behavior rather than to support a pooled solver ranking in the main manuscript. In this earlier Tier-2 analysis, SA and QPU emphasize different regions of attachment-distance space, with visibly different overlap profiles as well. This complements the main text by illustrating target dependence and backend-specific search behavior in discrete placement problems of this type.

\FloatBarrier

\subsection{QPU penalty-scale sweep}\label{app:note_s3}

\begin{figure}[!htbp]
  \centering
  \includegraphics[width=0.80\linewidth]{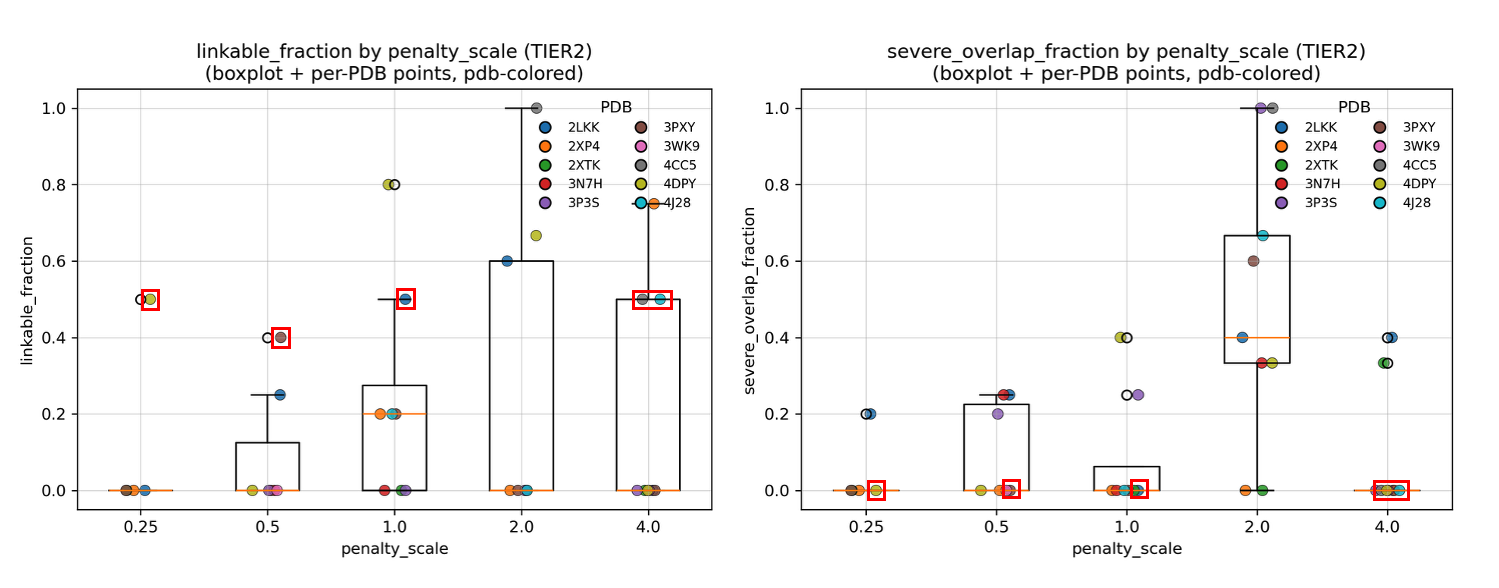}
  \caption{QPU penalty-scale sensitivity on the Tier-2 paired set. Linkable fraction and severe-overlap fraction vary substantially with penalty scale, supporting the interpretation that effective calibration is target dependent.}
  \label{fig:s4_penalty}
\end{figure}

Figure~\ref{fig:s4_penalty} examines penalty-scale sensitivity on the Tier-2 paired set. Both the linkable fraction and the severe-overlap fraction vary strongly with penalty scale, confirming that calibration is target dependent. This observation is consistent with the discussion in the main manuscript: a single fixed penalty regime is unlikely to transfer uniformly across diverse targets, especially when unary map energies and quadratic penalties interact differently across instances.

This sweep shows that performance differences in QPU-based optimization cannot be interpreted independently of calibration. It also helps explain why the additional backend analyses are presented as targeted studies of solver behavior rather than as blanket claims about universal solver rankings.

\FloatBarrier

\subsection{Hybrid versus QPU reference comparison}\label{app:note_s4}

\begin{figure}[!htbp]
  \centering
  \includegraphics[width=0.80\linewidth]{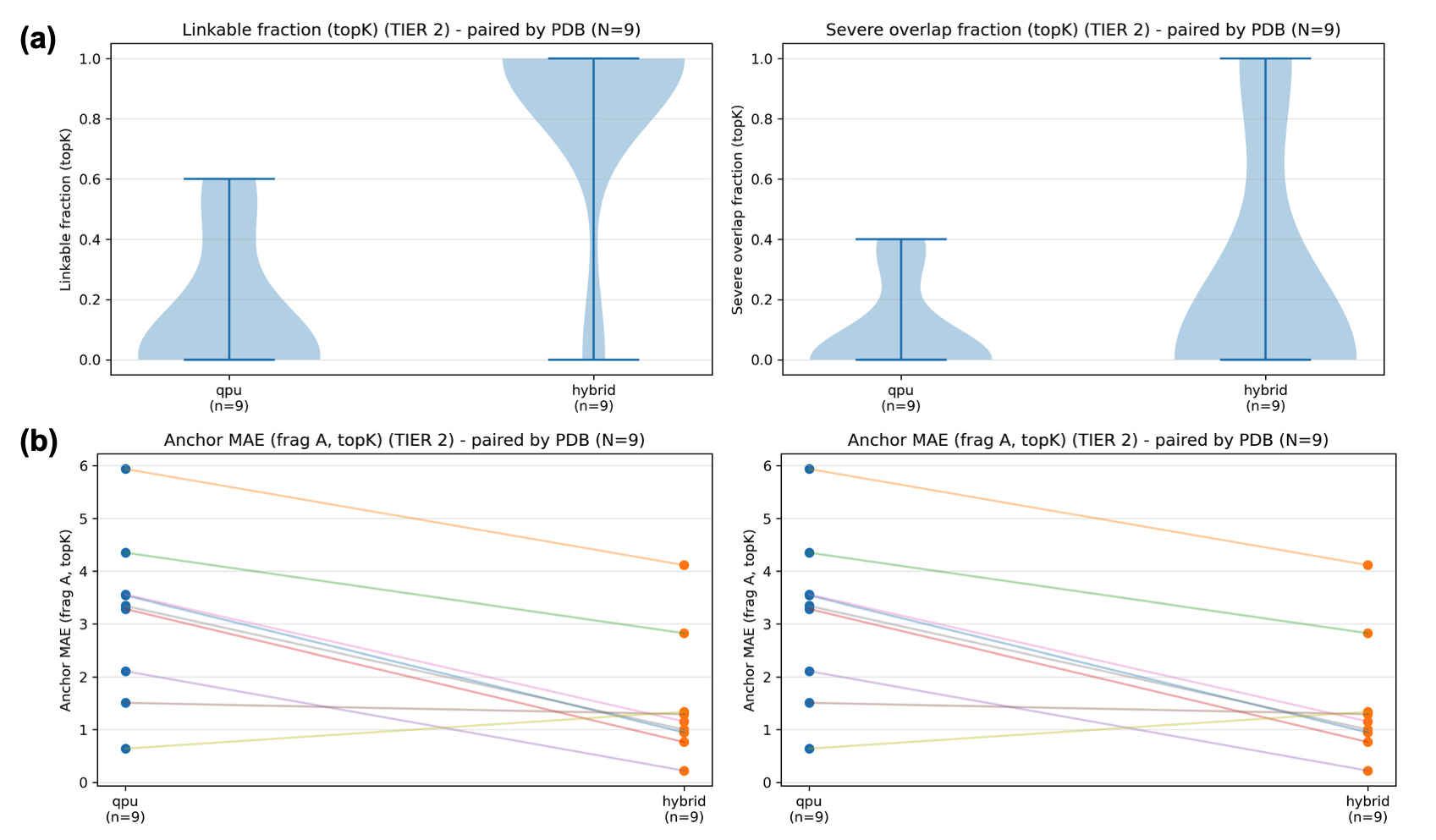}
  \caption{Paired hybrid-versus-QPU reference comparison on the selected Tier-2 set. Hybrid and direct QPU solvers exhibit different stability and diversity trade-offs, supporting the view that they serve as complementary solver options under limited quantum-computing resources.}
  \label{fig:s5_hybrid_qpu}
\end{figure}

Figure~\ref{fig:s5_hybrid_qpu} provides a paired reference comparison between hybrid and direct QPU optimization on the selected Tier-2 set. The point of this comparison is to show that they express different stability--diversity trade-offs under the same general optimization task.

In the paired comparison, the hybrid solver tends to return a more concentrated set of attachment-distance-feasible outcomes, whereas direct QPU sampling shows a broader spread of returned configurations. In practical use, the hybrid solver can therefore be attractive when robustness of returned feasible candidates is prioritized, whereas direct QPU sampling may be informative when diversity of returned configurations is of interest.

\FloatBarrier

\subsection{Supplementary hybrid follow-up on the stringent classical-failure subset}\label{app:note_s5}

The supplementary hybrid follow-up discussed in the main text was designed as a rescue analysis on the 259 targets that remained top-1 infeasible under both SA and CPLEX. This subset is deliberately stringent: it contains only cases that were not recovered by either the primary multi-read classical heuristic or the exact same-QUBO top-1 reference.

Under this subset analysis, the D-Wave Leap hybrid solver recovered feasible top-1 solutions in 125 of 259 cases, corresponding to 48.3\% of the stringent classical-failure subset. The corresponding no-distance-term hybrid control recovered feasible top-1 solutions in 57 of 259 cases, corresponding to 22.0\% of the same subset. This result is consistent with the main benchmark trend and shows that the rescue benefit under the hybrid backend remains strongly dependent on the inclusion of the inter-fragment distance term.

This subset-specific follow-up is the reason the hybrid solver appears in the main text as a rescue analysis rather than as part of the full matched benchmark.

\subsection{Rationale for hybrid use in the kinase case study}\label{app:note_s6}

The kinase case study in the main text uses the D-Wave Leap hybrid solver during the docking-stage QUBO optimization. This choice reflects the practical aim of the case study. The case study is intended to illustrate domain-guided deployment under a constrained candidate budget, not to serve as a matched solver benchmark against SA or CPLEX. In that setting, a hybrid backend provides a practical route for exploring the same formulation under a different search strategy while keeping the interpretation centered on the formulation and the domain priors.

\FloatBarrier
\subsection{Contextual docking-reference results and resource profiles}\label{app:tables}

To provide contextual references for the fragment-pair benchmark, we summarize docking-based reference results in Table~\ref{tab:contextual_baselines}. These values are included to show what happens when related placement tasks are handled by conventional docking workflows that do not explicitly optimize post-reconstruction linkability. Both rows correspond to the 775-case fragment-pair benchmark.

\begin{table}[!htbp]
\centering
\caption{Contextual docking-reference results for the fragment-pair benchmark.}
\label{tab:contextual_baselines}
\scriptsize
\setlength{\tabcolsep}{3pt}
\begin{tabular}{L{2.5cm} C{0.9cm} C{3.0cm} C{1.8cm} C{1.9cm} C{1.9cm}}
\toprule
Method & Targets & Top-1 definition & Top-1 feasible & Median RMSD$_1$ & Median RMSD$_2$ \\
\midrule
Vina individual & 775 & core rank 1 + back rank 1 & 15/775 (1.9\%) & 6.34 [2.87--16.43] & 6.19 [2.86--16.69] \\
Vina simultaneous & 775 & top-ranked co-docked pose & 0/775 (0.0\%) & 7.84 [3.26--17.14] & 8.61 [3.39--17.65] \\
\bottomrule
\end{tabular}
\end{table}
\FloatBarrier

Table~\ref{tab:resource_main} summarizes the logical resource profile of the main benchmark. Because the main-benchmark methods solve the same fixed-size QUBO instances, the table is intended primarily to provide a compact view of solver-time differences rather than differences in instance size.

\begin{table}[!htbp]
\centering
\caption{Resource profile of the main Q-SFD benchmark.}
\label{tab:resource_main}
\scriptsize
\setlength{\tabcolsep}{2pt}
\begin{tabular}{L{2.6cm} C{0.9cm} C{1.8cm} C{1.8cm} C{1.4cm} C{1.5cm} C{1.5cm}}
\toprule
Method & Targets & Top-1 feasible & Logical vars & \shortstack{$t_{\mathrm{build}}$\\(s)} & \shortstack{$t_{\mathrm{solve}}$\\(s)} & \shortstack{$t_{\mathrm{total}}$\\(s)} \\
\midrule
Q-SFD (SA) & 775 & 376/775 (48.5\%) & 170 [170--170] & 0.04 [0.03--0.04] & 3.05 [2.87--3.16] & 3.09 [2.91--3.20] \\
Q-SFD (SA, no distance term) & 775 & 183/775 (23.6\%) & 170 [170--170] & 0.04 [0.03--0.04] & 3.02 [2.86--3.14] & 3.06 [2.89--3.18] \\
Q-SFD (CPLEX) & 775 & 318/775 (41.0\%) & 170 [170--170] & 0.04 [0.03--0.04] & 0.06 [0.05--0.07] & 0.10 [0.09--0.11] \\
\bottomrule
\end{tabular}
\end{table}
\FloatBarrier

Table~\ref{tab:resource_quantum} summarizes the strict rescue-analysis subset for the supplementary D-Wave Leap hybrid follow-up.

\begin{table}[!htbp]
\centering
\caption{Resource profile of the supplementary D-Wave rescue analysis on the strict 259-target subset that remained top-1 infeasible under both Q-SFD~(SA) and CPLEX. Entries show per-target median [interquartile range]. Missing values indicate solver metadata not exposed in the recorded outputs.}
\label{tab:resource_quantum}
\scriptsize
\setlength{\tabcolsep}{2pt}
\begin{tabular}{L{3.3cm} C{0.9cm} C{1.8cm} C{1.8cm} C{1.8cm} C{1.8cm} C{1.5cm}}
\toprule
Method & Targets & Top-1 feasible & Logical vars & \shortstack{$t_{\mathrm{build}}$\\(s)} & \shortstack{$t_{\mathrm{solve}}$\\(s)} & \shortstack{$t_{\mathrm{total}}$\\(s)} \\
\midrule
Q-SFD (D-Wave Hybrid) & 259 & 125/259 (48.3\%) & 170 [170--170] & 0.03 [0.03--0.04] & 6.98 [6.88--7.21] & 7.01 [6.91--7.25] \\
Q-SFD (D-Wave Hybrid, no distance term) & 259 & 57/259 (22.0\%) & 170 [170--170] & 0.06 [0.06--0.07] & 6.99 [6.85--7.21] & 7.06 [6.92--7.27] \\
\bottomrule
\end{tabular}
\end{table}
\FloatBarrier

\end{appendices}
\clearpage
\bibliography{sn-bibliography}

\end{document}